\documentclass[onecolumn]{article}

 \evensidemargin \oddsidemargin

\usepackage{graphicx}          %
\usepackage{bm}%
\usepackage{amsmath,amssymb}%
\usepackage{natbib}
\usepackage[skip=3mm]{parskip}
\usepackage[normalem]{ulem}
\title{Interval Analysis of Worst-case Stationary Moments for Stochastic Chemical Reactions with Uncertain Parameters}
\author{Yuta Sakurai, Yutaka Hori%
\thanks{Y. Sakurai and Y. Hori are with Department of Applied Physics and Physico-Informatics, Keio University \newline
3-14-1 Hiyoshi, Kohoku-ku, Yokohama, Kanagawa 223-8522, Japan. \newline
Corresponding author Y.~Hori. Tel. +81-45-566-1548.}}%

\date{}          

\makeatletter
\def\thanks#1{\protected@xdef\@thanks{\@thanks
\protect\footnotetext{#1}}}
\makeatother

 \newcommand{\PP}{\mathbb{P}}
\begin{document}
\maketitle

\begin{abstract}
The dynamics of cellular chemical reactions are variable due to stochastic noise from intrinsic and extrinsic sources. The intrinsic noise is the intracellular fluctuations of molecular copy numbers caused by the probabilistic encounter of molecules and is modeled by the chemical master equation. The extrinsic noise, on the other hand, represents the intercellular variation of the kinetic parameters due to the variation of global factors affecting gene expression. 
The objective of this paper is to propose a theoretical framework to analyze the combined effect of the intrinsic and the extrinsic noise modeled by the chemical master equation with uncertain parameters. More specifically, we formulate a semidefinite program to compute the intervals of the stationary solution of uncertain moment equations whose parameters are given only partially in the form of the statistics of their distributions. 
The semidefinite program is derived without approximating the governing equation in contrast with many existing approaches. Thus, we can obtain guaranteed intervals of the worst possible values of the moments for all parameter distributions satisfying the given statistics, which are prohibitively hard to estimate from sample-path simulations since sampling from all possible uncertain distributions is difficult.
We demonstrate the proposed optimization approach using two examples of stochastic chemical reactions and show that the solution of the optimization problem gives informative upper and lower bounds of the statistics of the stationary copy number distributions.

\vspace*{-3mm}
\end{abstract}


\section{Introduction}
\label{intro-sec}
The stochastic response of biomolecular reactions in cells is often explained by two types of noise called intrinsic and extrinsic noise \citep{Elowitz2002, Taniguchi2010}. 
The intrinsic noise is the intracellular fluctuations of molecular copy numbers caused by the probabilistic encounter of molecular species such as mRNA and proteins in a single cell. %
The extrinsic noise, on the other hand, arises from the intercellular variation of the global factors affecting gene expression, and %
some of these are modeled by the variation of the rate parameters of the reactions. %

\par
The dynamics of the intrinsic noise is modeled by a continuous-time discrete state Markov process on a possibly infinite integer lattice associated with the copy numbers of molecular species, whose governing equation is called the chemical master equation (CME) \citep{McQuarrie1967, Gillespie1992}. %
However, the exact solution of the CME is hard to obtain since the number of the states, which is equal to the order of the equation, becomes extremely large or even infinite in applications of practical interest. %
Thus, analyses of stochastic chemical reactions are carried out either by sample-path generation using the stochastic simulation algorithm \citep{Gillespie1976} or by approximate models.

\par
Examples of the approximate models include the chemical Langevin equation \citep{Gillespie2000}, the linear noise approximation  \citep{vanKampen2007}, and the truncated moment equations \citep{Singh2011, Lakatos2015, Schnoerr2015}, which allow for computing approximate sample paths or dynamic moments of the molecular copy numbers of interest. 
Efforts were also made to  %
theoretically guarantee the accuracy of analysis by bounding the error of the approximation.
For instance, \cite{Munsky2006, Gupta2017} proposed the finite state projection, which enables analytic quantification of the error bound of the copy number distributions. \cite{Ahmadi2016} developed a method for bounding the solution of the Langevin equation.
More recently, \cite{Ghusinga2017, SakuraiCDC2017, SakuraiRSInt2018, SakuraiLCSS2019, Dowdy2018, DowdyTransient2018, Kuntz2019} 
 independently proposed an optimization based approach for bounding the solution of truncated moment equations based on the SDP relaxation of the generalized moment problem \citep{Lasserre2009}, of which the idea was extended to the analysis of a wider class of systems \citep{Lamperski2017, Lamperski2019, Ghusinga2020}.

\par
Despite these advancements, one limitation of these general frameworks is that they focus only on the analysis of intrinsic noise while experimental observations suggest that the stochastic cellular response is the result of the combined effects of intrinsic and extrinsic noise \citep{Taniguchi2010}. 
Thus, an important next step is to generalize these frameworks to enable simultaneous analysis for intrinsic and extrinsic noise.
\par
\begin{figure}[t]
 \centering
 \includegraphics[width=11cm]{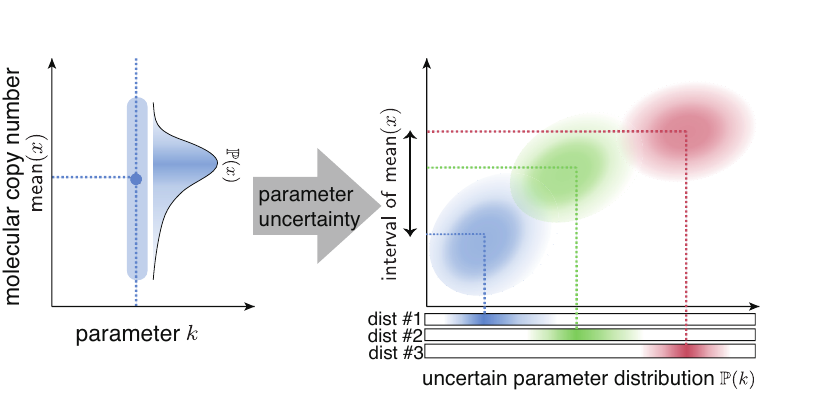}
\caption{The concept of the worst-case interval of stationary moments for uncertain parameter distributions}
\label{fig:ConceptFigure}
 \end{figure}

Toward this goal, this paper considers a computational method to obtain theoretically guaranteed bounds of the stationary moments of the copy number distributions subject to extrinsic noise modeled by the uncertainty of reaction rates.
Since exact identification of the uncertainty is hard in practice, we here assume that only part of the statistics of the parameter distribution such as the mean is available. %
This implies that the stationary moments of the copy number distribution can be obtained only as the {\it worst-case} interval for all possible parameter distributions satisfying the {\it a priori} statistics (Fig. \ref{fig:ConceptFigure}). 
We show that the problem of the worst-case interval analysis reduces to a similar form of the semidefinite program that was designed for the deterministic parameter case \citep{Ghusinga2017, SakuraiCDC2017, SakuraiRSInt2018,Dowdy2018, Kuntz2019} by reorganizing the CME and adding various types of constraints to the optimization problem.
In particular, we show that the proposed optimization program is capable of computing informative bounds on the stationary moments of highly uncertain moment equations that are hard to analyze with the widely-used stochastic simulation algorithm \citep{Gillespie1976}.

\par
The organization of this paper is as follows. In Section \ref{model-sec}, we formally address the worst-case analysis problem to be solved. In Section \ref{momeq-sec}, we formulate the optimization problem for computing valid bounds of uncertain stationary moments. Then, specific forms of the optimization constraints for characterizing the set of uncertain parameter distributions are presented in Section \ref{res-sec}. Section \ref{appl-sec} is devoted to the demonstration of the proposed approach using two illustrative examples. Finally, we summarize the results in Section \ref{concl-sec}.

\par
{\bf  Notations:}
$\mathbb{N}_0$ is the set of natural numbers including zero, $\mathbb{N}_0:=\{0,1,2,\ldots\}$,  $\mathbb{Z}$ is the set of integers, and $\mathbb{R}_{>0}$ is the set of positive real numbers, $\mathbb{R}_{> 0}:=\{x\in\mathbb{R}~|~x > 0\}$. A superscript is used to represent the dimension of the vector space, {\it e.g.,} $\mathbb{N}_0^n$.
A probability distribution defined on the sample space $X$ and its support is denoted by $\mathbb{P}_X$ and ${\rm supp}(\mathbb{P}_X)$, respectively. The probability $\mathbb{P}_X(\bm{X} = \bm{x})$ is denoted by $\mathbb{P}_X(\bm{x})$, and, when necessary, time $t$ is explicitly displayed as $\mathbb{P}_X(\bm{x}; t)$.
A scalar $\bm{X}^{\bm \alpha}$ is defined for vectors
$\bm{X} = [X_1, X_2, \ldots, X_n]^\top$ and $\bm{\alpha} = [\alpha_1, \alpha_2, \ldots, \alpha_n]^\top$ 
as %
$\bm{X}^{{\bm{\alpha}}}:=\prod_{i=1}^n X_i^{\alpha_i}= X_1^{\alpha_1}X_2^{\alpha_2}\ldots X_n^{\alpha_n}$.
$\mathbb{E}[\bm {X}^{\bm{\gamma}}]$ with $\|\bm{\gamma} \|_1 = p$ denotes a $p$-th order moment of  $\mathbb{P}_{{X}}$ defined by  
\begin{align}
\mathbb{E}[{\bm {X}}^{\bm{\gamma}}] := \int_{ \mathcal{X}}  \bm{x}^{\bm{\gamma}} \ \mathrm{d}\mathbb{P}_{X}({\bm{x}}), \notag
\end{align}
where $\mathcal{{X}} := \mathrm{supp}(\mathbb{P}_{{X}})$.

 \section{Model of Uncertain Stochastic Chemical Reactions and Problem Formulation}
 \label{model-sec}
In this section, we first introduce a chemical master equation, a mathematical model of stochastic chemical reactions, and define the problem of moment analysis with uncertain reaction parameters. 
\par
Consider a chemical reaction system that consists of $n$ species of molecules, $\mathcal{M}_1, \mathcal{M}_2, \ldots, \mathcal{M}_n$, and $r$ reactions. 
We denote the copy number of the molecular species $\mathcal{M}_j$ by $X_j$ and define $\bm{X}:=[X_1,X_2\ldots,X_n]^\top\in \mathbb{N}_0^n$. The stoichiometry of the $i$-th reaction is defined by $\bm{s}_i \in\mathbb{Z}^n$, meaning that the molecular copy numbers change from ${\bm X}$ to ${\bm X} + {\bm s}_i$ by reaction $i$. The reactions occur in a stochastic manner due to the low copy nature of the molecular species, and thus, the dynamics of the copy number $\bm{X}$ is considered as a stochastic process. More specifically, the probability of the occurrence of reaction $i$ in an infinitesimal time $dt$ is given by $w_i({\bm X}, K_i) dt$, where $w_i({\bm X}, K_i)$ is the propensity function with a constant $K_i \in \mathbb{R}_{> 0}$.
We assume that all reactions are elementary, meaning that the propensity function is a zero-th, first, or second order polynomial in $X_i$ $(i=1,2,\ldots,n)$. %

\par
Let $\mathbb{P}_{X|K,X_0}$ denote the conditional probability distribution of $\bm{X}$ given 
the time-invariant rate constants ${\bm K} := [K_1, K_2, \ldots, K_r]^\top \mathbb{R}^r_{> 0}$ and the initial value of the copy number ${\bm X}_0 \in \mathbb{N}_0^n$.

The evolution of the distribution is then governed by the chemical master equation (CME) 
\citep{Gillespie1992}
\begin{align}
\frac{d}{dt}&\mathbb{P}_{X|K,X_0}(\bm{x}|\bm{k},\bm{x}_0; t) \notag \\
&=\sum_{i=1}^r\{w_i(\bm{x}-\bm{s}_i,k_i) \mathbb{P}_{X|K,X_0}(\bm{x}-\bm{s}_i|\bm{k},\bm{x}_0; t)\notag \\
 & -w_i(\bm{x},k_i)\mathbb{P}_{X|K,X_0}(\bm{x}|\bm{k},\bm{x}_0; t) \}. \label{CME}
\end{align}
The CME is also known as Kolmogorov's forward equation for a discrete state Markov chain, where the state of the chain is the copy number of molecular species.

\par
The CME (\ref{CME}) characterizes the dynamics of the intrinsic variability caused by the stochastic reaction events within a cell. 
On the other hand, the cell population is also subject to extrinsic noise resulting from the variation of global factors. %
Hence, we here consider the extrinsic noise that can be modeled by the variation of the time-invariant kinetic parameters across the cell population that is characterized by a distribution $\mathbb{P}_K$.

\par
In what follows, we consider analyzing the stationary moments of the copy number distribution ${\PP}_{X}$ when the parameter distribution $\mathbb{P}_K$ is partially given in the form of its moments. 
Specifically, our goal is to propose a mathematical optimization program for computing valid bounds
of the stationary moments and their associated statistical values such as the mean and the variance of the distribution based on the {\it a priori} information of the parameter distribution $\mathbb{P}_K$.
More formally, the problem is stated as follows.

\par
\noindent
{\bf Problem.}\ Consider the chemical master equation (\ref{CME}). %
Suppose a set of parameter distributions $\mathcal{P}$ is given. Compute mathematically valid upper and lower bounds of the stationary moments of $\mathbb{P}_X$ for all parameter distributions $\mathbb{P}_K \in \mathcal{P}$ and all initial distributions $\mathbb{P}_{X_0}$.

\par
It is reasonable, in practice, to assume that the actual distribution of the parameters $\mathbb{P}_K$ is unknown but only some statistics such as the mean and the covariance are known. %
Thus, we here consider the case where the set $\mathcal{P}$ is characterized by some of the moments of parameter distributions. 
The stationary distribution of the copy numbers ${\mathbb{P}}_X$ might not be unique for the set of the parameter distributions $\mathcal{P}$. 
In other words, we can obtain only an interval of statistics of the copy number distribution. The computed upper and lower bounds of the statistics then gives a valid range of the worst-case statistics for the stochastic chemical system (\ref{CME}) when the underlying parameter distribution $\mathbb{P}_K \in \mathcal{P}$ is uncertain (Fig. \ref{fig:ConceptFigure}). 
In what follows, we impose the following assumptions to enable moment based analysis of the stationary distribution of the molecular copy numbers.

 \noindent
{\bf Assumption 1.}\ 
For any parameter distributions $\mathbb{P}_K$ in the given set $\mathcal{P}$, and any initial copy number distributions $\mathbb{P}_{X_0}$, 
(i) the stationary solution of the CME (\ref{CME}) exists, and (ii) its associated Markov chain is non-explosive. Moreover, (iii) all moments of the stationary distributions $\mathbb{P}_{X}$ are finite.

\noindent
{\bf Remark 1.} 
The conditions (i) and (ii) guarantee the existence of the stationary distributions (Theorem 30 in \cite{Kuntz2019}). The condition (iii) is necessary to rule out the case of heavy-tailed copy number distributions as observed in \cite{Ham2020}, in which case moment based characterization of the stationary distribution is not possible.

\section{Mathematical Optimization for the Worst-case Analysis}
\label{momeq-sec}
To analyze the uncertain stationary moments of the copy number distribution, we first introduce the moment equation of the joint distribution of the molecular copy number $\bm{X}$ and the parameter $\bm{K}$. %
To this goal, we reorganize the CME (\ref{CME}) by marginalizing the parameter $\bm{K}$ and the initial copy number $\bm{X}_0$, and incorporating the parameter into the state by $\hat{\bm{X}} := [\bm{X}^\top, \bm{K}^\top]^\top$. Specifically, eq. (\ref{CME}) becomes 
\begin{align}
\frac{d}{dt}\mathbb{P}_{\hat{X}}(\hat{\bm{x}}; t)=\sum_{i=1}^r\{&\hat{w}_i(\hat{\bm{x}}-\hat{\bm{s}}_i)  \mathbb{P}_{\hat{X}}(\hat{\bm{x}}-\hat{\bm{s}}_i; t) \notag \\
& -\hat{w}_i(\hat{\bm{x}}) \mathbb{P}_{\hat{X}}(\hat{\bm{x}}; t) \},
\label{newCME-eq}
\end{align}
where 
$\hat{\bm{s}}_i := [\bm{s}_i^\top,\ \bm{0}^\top]^\top$ and $\hat{w}_i(\bm{\hat{x}}) := w_i(\bm{x}, k_i)$.

\par
Eq. (\ref{newCME-eq}) can be viewed as a chemical master equation for the new state $\hat{\bm{X}}$. In particular, the rate constants %
are incorporated into the state. %
Thus, the dynamics of the moments of the distribution $\mathbb{P}_{\hat{X}}$ can be modeled by the moment equation using the standard approach (see \cite{SakuraiRSInt2018} for example). 
This allows us to recast the analysis problem of the uncertain stationary moments into an optimization problem that was previously studied for computing valid moment bounds of stochastic reactions without parameter uncertainty \citep{Ghusinga2017, SakuraiCDC2017, SakuraiRSInt2018,Dowdy2018, Kuntz2019}.

\par

The moment equation is a set of linear ordinary differential equations of the moments of $\mathbb{P}_{\hat{X}}$, and its stationary solution gives the stationary moment. %
The stationary moment equation is specifically given by 
\begin{align}
0=\sum_{i=1}^r \sum_{\bm{\gamma}} a^{\bm{\zeta}}_{i,\bm{\gamma}} \mathbb{E}[\hat{\bm{X}}^{\bm{\gamma}}]\label{parmomeq}
\end{align}
for each $\bm{\zeta} := [\zeta_1, \zeta_2, \ldots, \zeta_{n+r}] \in \mathbb{N}_0^{n+r}$, where %
the constant $a^{\bm{\zeta}}_{i, \bm{\gamma}}$ is the coefficient of $\hat{\bm{X}}^{\bm{\gamma}}$ in the polynomial $\{(\hat{\bm{X}}+\hat{\bm{s}}_i)^{\bm{\zeta}}-\hat{\bm{X}}^{\bm{\zeta}}\}\hat{w}_i(\hat{\bm{X}})~(i=1,2,\ldots,r)$ and $\bm{\gamma} \in \mathbb{N}_0^n$ is the exponent (see Notations in Section \ref{intro-sec}). 
A finite subset of eq. (\ref{parmomeq}) can then be written as 
\begin{align}
\bm{0} = A\bm{\mu} + B\bm{\nu} + C\bm{\xi}, \label{parmomeqst}
\end{align}
where %
\begin{align}
&\bm{\mu} := [{\mathbb{E}}[\bm{X}^{\bm{\alpha}_1}], {\mathbb{E}}[\bm{X}^{\bm{\alpha}_2}], \ldots, {\mathbb{E}}[\bm{X}^{\bm{\alpha}_{n_1}}]]^\top, \notag \\ %
&\bm{\nu} := [{\mathbb{E}}[\bm{X}^{\bm{\alpha}_1}\bm{K}^{\bm{\beta}_1}], {\mathbb{E}}[\bm{X}^{\bm{\alpha}_1}\bm{K}^{\bm{\beta}_2}], \ldots, 
{\mathbb{E}}[\bm{X}^{\bm{\alpha}_{n_2}}\bm{K}^{\bm{\beta}_{n_3}}]
]^\top, \notag \\
&\bm{\xi} := [{\mathbb{E}}[\bm{K}^{\bm{\beta}_1}], {\mathbb{E}}[\bm{K}^{\bm{\beta}_2}], \ldots, \mathbb{E}[\bm{K}^{\bm{\beta}_{n_4}}]^\top  \notag 
\end{align}
are finite dimensional vectors of moments obtained by truncating all moments beyond a chosen order, and the matrices $A, B$, and $C$ are defined with the coefficients in eq. (\ref{parmomeq}). 

Eq. (\ref{parmomeqst}) implies that, given the stoichiometry $\hat{\bm{s}}$ and the propensity $\hat{w}_i(\hat{\bm{X}})$, the stationary moment of the copy number distribution $\mathbb{P}_{X}$ can be found as a solution of the linear equation 
\begin{align}
A \bar{\bm{\mu}} + B \bar{\bm{\nu}} + C \bar{\bm{\xi}} = 0, 
\label{linear-eq}
\end{align}
where $\bar{\bm{\mu}}$, $\bar{\bm{\nu}}$ and $\bar{\bm{\xi}}$ are vectors of independent variables. 
When the information of the parameter distribution $\mathbb{P}_K$ is partially known in the form of moments, the variable $\bar{\bm{\xi}}$ is constrained by the {\it a priori} information of the moments $\Omega$, {\it i.e.,} $\bar{\bm{\xi}} \in \Omega$.
This means that the stationary moment of the copy number distribution can be found as a solution of the linear equation (\ref{linear-eq}) subject to the constraint $\bar{\bm{\xi}} \in \Omega$, which will be discussed in detail in Section \ref{res-sec}. 
In general, however, the linear equation is highly underdetermined, {\it i.e.,} there is no subset of equations that can close the system of equations,
and simply solving the linear equation (\ref{linear-eq}) does not give informative moment bounds.

\par

Hence, we here introduce additional necessary conditions that the solution of the linear equation (\ref{linear-eq}) must satisfy, and formulate an optimization problem that bounds the target moment of the copy number distribution. 
\noindent
{\bf Proposition 1.}\ 
{\it 
Consider the stochastic reaction system governed by the CME (\ref{CME}). 
Let $\mathcal{P}$ denote a given set of uncertain parameter distributions characterized by the constraints of their moments $\Omega$, {\it i.e.,} %
$\mathcal{P} := \{\mathbb{P}_K\ |\ \bm{\xi} \in \Omega\}$. 
Suppose Assumption 1 holds, and polynomials $c_i(\bm{K})$ and $d_i(\bm{X})$ satisfying 
\begin{align}
&\mathrm{supp}({\mathbb{P}}_K) \subseteq \{\bm{K}\ |\ c_{i}(\bm{K}) \ge 0\}_{i=1}^{\ell_1}
 \notag \\
&\mathrm{supp}({\mathbb{P}}_X) \subseteq \{\bm{X}\ |\ d_{i}(\bm{X}) \ge 0\}_{i=1}^{\ell_2}. \notag 
\end{align}
are given. 
Let $\varphi^*_{\min}$ (resp., $\varphi^*_{\max}$) denote the minimum (resp., maximum) value of the stationary moment $\bm{f}^\top \bm{\mu}$  among all possible  $\mathbb{P}_K \in \mathcal{P}$, {\it i.e.,} $\varphi^*_{\min} := \min_{\mathcal{P}} \bm{f}^\top \bm{\mu}$ (resp., $\varphi^*_{\max} := \max_{\mathcal{P}} \bm{f}^\top \bm{\mu}$), where $\bm{f}$ is a given constant vector for defining the moments of interest. 
Then, the solution of the following minimization (resp., maximization) problem gives the lower (resp., upper) bound of $\varphi^*_{\min}$.
\begin{equation}
\begin{aligned}
&\min_{\bar{\bm{\mu}}, \bar{\bm{\nu}}, \bar{\bm{\xi}}}\   \bm{f}^\top \bar{\bm{\mu}}     \ \ \mathrm{subject\ to\ } \   \\
&A \bar{\bm{\mu}} + B \bar{\bm{\nu}} + C \bar{\bm{\xi}} = 0, \ \bar{\bm{\xi}} \in \Omega \\
&
\bar{H}_i \!:=\!
 \begin{cases}
 \mathcal{L}(\mathbb{E}[\bm{g}_0 \bm{g}_0^\top]) \! \succeq \! O \!\!\!&\!\! (i=0)\\
 \mathcal{L}(\mathbb{E}[{c}_i(\bm{K}) \bm{g}_i \bm{g}_i^\top]) \! \succeq \!O \!\!\!&\!\!  (i=1,\ldots,\ell_1) \\ 
 \mathcal{L}(\mathbb{E}[{d}_{i-\ell_1}(\bm{X}) \bm{g}_i \bm{g}_i^\top]) \! \succeq \! O \!\!\!&\!\!  (i=\ell_1\!+\!1, \ldots,\ell_1\!+\!\ell_2), %
\end{cases}
\end{aligned}
\notag
\end{equation}
where 
$\bar{\bm{\mu}} := \mathcal{L}(\bm{\mu})$, 
$\bar{\bm{\nu}} := \mathcal{L}(\bm{\nu})$, and 
$\bar{\bm{\xi}} := \mathcal{L}(\bm{\xi})$ with  
a bijective operator $\mathcal{L}(\cdot)$  that maps each moment in the entries to an independent variable, and 
$\bm{g}_i$ is a vector of the monomial basis of polynomials $\mathbb{R}[\bm{ \hat{X} }]$ of an arbitrary degree.}

The linear matrix inequality (LMI) conditions for the matrices $\bar{H}_i \ (i=0,1,2,\ldots,\ell_1 + \ell_2)$ are the necessary conditions for the variables to be the moments of the joint distribution $\mathbb{P}_{\hat{X}}$ since the entries of $\bar{H}_i$ correspond to the moments of the joint distribution. 
Thus, the optimization problem explores valid bounds of uncentered moments $\bm{f}^\top \bm{\mu}$ over the set of variables that the moments of $\mathbb{P}_{\hat{X}}$ must satisfy. 

In particular, 
if the constraints $\Omega$ on the moments of $\mathbb{P}_K$ is represented by LMIs, the optimization problem becomes a semidefinite program (SDP), which will be discussed in detail in Section \ref{res-sec}. 
More specifically, this class of optimization is known as an SDP relaxation of the generalized moment problem \citep{Lasserre2009}. %
The use of such SDP relaxation was previously studied for computing valid bounds of the moments when the parameter $K_i$ of the propensity function $w_i(\bm{X}, K_i)$ is given and deterministic \citep{Ghusinga2017, SakuraiCDC2017, SakuraiRSInt2018,Dowdy2018, Kuntz2019}. Proposition 1 extends these results to enable the exploration of the {\it worst-case} moments when the parameters are given as an uncertain set of distributions $\mathcal{P}$.
\noindent
{\bf Remark 2.}
When a conservation law holds for some molecular species, the redundant state variables in the CME can be systematically removed by using the bases of the left null space of the stoichiometry matrix $[\bm{s}_1, \bm{s}_2, \ldots, \bm{s}_r]$ as shown in \cite{Dowdy2018}. This allows for reducing the number of variables and tightening computed bounds of the optimization problem.

\section{Formulation of Uncertain Parameter Set for Optimization}
\label{res-sec}

\par
In this section, we introduce specific forms of the constraints $\Omega$ in the optimization problem in Proposition 1 by considering typical analysis problems of stochastic biomolecular reactions. 
In particular, we show 
the constraints that arise in many practical analysis problems can be expressed by linear (matrix) inequalities, enabling the optimization problem to be computed by SDP solvers. 

\subsection{Uncertain parameter distributions in practical analysis}

\par
In practice, the joint distribution $\mathbb{P}_K$ of the rate parameters is rarely identified from experimental data because of the sparse measurement and the lack of well-established methodology. Instead, the rate parameters are only expressed as the mean value $\mu_i$ and the standard deviation $\sigma_i$ of the  marginal distribution of the parameter $\mathbb{P}_{K_i}$, which are defined by 
\begin{align}
& \mu_i:=\mathbb{E}[K_i], %
\ \sigma_i^2:=\mathbb{E}[K_i^2]-(\mathbb{E}[K_i])^2. 
\label{mean-variance-eq}
\end{align}
The correlation between the parameters 
\begin{align}
\mathrm{corr}(K_i, K_j) := \frac{\mathbb{E}[K_iK_j]-\mu_i\mu_j}{\sqrt{\mathbb{E}[K_i^2]-\mu_i^2} \sqrt{\mathbb{E}[K_j^2]-\mu_j^2}} %
\label{corr-eq}
\end{align}
 is also an important factor that is identified or estimated from experimental data since, in a single cell,  the rate parameters are affected by shared resource molecules and common environmental factors such as ribosomes, RNA polymerases, and temperature \citep{Boo2019,Taniguchi2010}.

\par
Therefore, for the worst-case analysis, the set of parameter distributions $\mathcal{P}$ needs to be explored based on the partial information of (i) the mean, (ii) the variance, and (iii) the correlation between the parameters. Moreover, these statistics themselves are potentially uncertain due to the limitation of parameter identification.
In what follows, we consider specific forms of the constraints $\Omega$ to solve these analysis problems. %

\par

\subsection{Worst-case analysis with known moment values}

\par
Let us first consider the case where the mean and the variance (\ref{mean-variance-eq}) of the uncertain parameter distribution are given. 
In literatures, the marginal distribution of each parameter is often assumed to be a parametric distribution such as a gamma distribution (see \cite{Taniguchi2010}, for example). 
Then, the moments of the marginal distributions $\mathbb{P}_{K_i}$ can be analytically obtained, and 
$\bar{\bm{\xi}}$ in Proposition 1 can simply be set to the values computed by the analytic solution.
For example, the moments of gamma distributions are given by
\begin{align}
\mathbb{E}[K_i^{\beta_i}]=\theta_i^{\beta_i}\prod_{j=1}^{\beta_i}(\eta_i+j-1),  \label{mom_gamma}
\end{align}
using the two parameters $\theta_i$ and $\eta_i$, which can be identified from the given mean and variance.

\par
In reality, however, the parameter distribution is not necessarily parametric. Thus, it is more reasonable to explore all possible parameter distributions $\mathbb{P}_{K}$, {\it i.e.} the set of distributions $\mathcal{P}$, constrained only by the first and the second order moments of the marginal distributions $\mathbb{P}_{K_i}$. 
The following corollary summarizes the constraints $\Omega$ of the optimization problem in Proposition 1 when part of the statistics in eqs. (\ref{mean-variance-eq}) and (\ref{corr-eq}) is given. %
\noindent
\textbf{Corollary 1.}
{\it 
Suppose the mean $\mu_i$ and the variance $\sigma_i^2$ of the marginal distribution $\mathbb{P}_{K_i}$ are given, and the bound of the correlation between the parameters is given by $|\mathrm{corr}(K_i, K_j)| \le r_{ij}$. 
Let the constraints $\Omega$ be set as 
\begin{equation}
\left\{
\begin{aligned}
& \mathcal{L}(\mathbb{E}[K_i]) - \mu_i = 0, \\ 
& \mathcal{L}(\mathbb{E}[K_i^2]) - \mu_i^2 - \sigma_i^2 = 0, \\ 
&-\mathcal{L}(\mathbb{E}[K_i K_j])+\mu_i\mu_j+r_{ij}\sigma_i\sigma_j\ge0, \\
& \mathcal{L}(\mathbb{E}[K_i K_j]) - \mu_i \mu_j+r_{ij}\sigma_i\sigma_j\ge 0.
\end{aligned}
\right.
\label{cor1-eq}
\end{equation}
Then, the solution of the minimization problem in Proposition 1 gives a valid lower bound of $\bm{f}^\top \bm{\mu}$.
}

\par

\subsection{Worst-case analysis with uncertain moment values}
When the number of experimental data is not sufficient, the variance itself could also be uncertain and is given as an interval by $\underline{\sigma}_i^2 \le \sigma_i^2 \le \overline{\sigma}_{i}^2$. 
As shown in the next Corollary, the optimization problem in Proposition 1 can also incorporate such uncertainty as semidefinite constraints, allowing for the problem to be solved by SDP solvers.

\textbf{Corollary 2.}
{\it 
Suppose the mean values $\mu_\ell$ of the marginal distributions $\mathbb{P}_{K_\ell}\ (\ell=i,j)$ are given, and the interval of the variance and the correlation between the parameters are given by  $0 < \underline{\sigma}_i^2 \le \sigma_i^2 \le \overline{\sigma}_i^2$ and  $0 < |\mathrm{corr}(K_i, K_j)| \le r_{ij}$, respectively.  %
Let the constraints $\Omega$ be set as 
\begin{equation}
\left\{
\begin{aligned}
& 
\underline{\sigma}_\ell^2 \le \mathcal{L}(\mathbb{E}[K_{\ell}^2])-\mu_\ell^2 \le \overline{\sigma_\ell}^2 \ \ (\ell = i, j), \notag \\
&\left[\begin{array}{cc}
r_{ij}(\mathcal{L}(\mathbb{E}[K_i^2])-\mu_i^2)&\mathcal{L}(\mathbb{E}[K_iK_j])-\mu_i\mu_j\\
\mathcal{L}(\mathbb{E}[K_i K_j])-\mu_i\mu_j&r_{ij}(\mathcal{L}(\mathbb{E}[K_j^2])-\mu_j^2)
\end{array}\right]\succeq O. \label{mom_corr}
\end{aligned}
\right.
\end{equation}
Then, the solution of the minimization problem in Proposition 1 gives a valid lower bound of $\bm{f}^T \bm{u}$.
}%

\par

In Corollary 2, the semidefinite constraint is obtained by the Schur complement of the inequality $(\mathrm{corr}(K_i, K_j))^2 \le r_{ij}^2$. 
In the case of $r_{ij}=0$, the parameters are not correlated, {\it i.e.} $\mathbb{E}[K_iK_j]=\mu_i\mu_j$, and thus, the cross-moment can be immediately substituted into $\mathbb{E}[\bar{\bm{\xi}}]$.

\par
It should be noted that the proposed optimization can flexibly incorporate the information of parameter distributions $\mathbb{P}_K$. In Corollaries 1 and 2, the only constraints on $\mathbb{P}_{K}$ are the partial statistics of the distribution, and no parametric distributions need to be assumed. However, if necessary, one can also assume parametric distributions and can easily incorporate the associated constraints into the proposed optimization framework, as demonstrated in the next section. 

\section{Application examples}
\label{appl-sec}
In this section, we demonstrate the proposed optimization approach by using two examples of stochastic chemical reactions and show that the solution of the optimization problem gives informative upper and lower bounds of the statistics of the stationary copy number distribution $\PP_X$.
\begin{table}[t]
\centering
\caption{Specification of dimerization process (\ref{ModelMonomerDimer})}
\label{tbl:ParamMonomerDimer}
\begin{tabular}{ccl} \hline
index $i$ & ~~~Reaction rate $w_i$~~~ & ~~~Stoichiometry $\bm{s}_i$~~~ \\ \hline
1 & $w_1=K_1 D$ & ~~~$s_1=1$ \\
2 & $w_2=K_2 X$ & ~~~$s_2=-1$ \\
3 & $w_3=K_3 X(X-1)$ & ~~~$s_3=-2$ \\ \hline
\end{tabular}
\end{table}

\subsection{Analysis for partially known parameter distributions}
\label{partial-example}
We consider the dimerization process of a molecular species A, which consists of three reactions: 
\begin{align}
\phi\xrightarrow{w_1} A,~~~A\xrightarrow{w_2} \phi,~~~2A\xrightarrow{w_3} A:A,\label{ModelMonomerDimer}
\end{align}
where the copy number of the molecular species A is denoted by $X$, and the stoichiometry and the propensity functions are defined in Table \ref{tbl:ParamMonomerDimer}. These reactions correspond to gene expression, degradation, and protein dimerization, for instance.
Suppose the mean and the variance of the marginal parameter distribution $\mathbb{P}_{K_1}$ are 0.8 and 0.32, respectively, and those of  $\mathbb{P}_{K_2}$ are 0.4 and 0.04, respectively.
We assume that these marginal distributions are gamma distributions. Then, the shape factor $\eta_i$ and the scale factor $\theta_i$ for $K_i$ in eq. (\ref{mom_gamma}) are identified as $(\eta_1, \theta_1)=(2,0.4)$ and $(\eta_2,\theta_2)=(4,0.1)$, respectively. $K_3$ and $D$ are assumed to be constants, and $K_3 = 0.02$ and $D = 5$.
The parameters $K_1$ and $K_2$ are possibly correlated and the correlation is bounded by $|\mathrm{corr}(K_1, K_2)| \le r$.
This constraint is expressed by the linear inequality conditions as shown in eq. (\ref{cor1-eq}).

In what follows, we analyze the stationary mean copy number of the monomer $A$. To this goal, the moment equation (\ref{parmomeqst}) is computed based on the CME (\ref{CME}). 
For instance, 
\begin{align}
&\bm{0}=\left[
\begin{array}{cc}
2 K_3&-2 K_3\\
0&0\\
0&0
\end{array}\right]\!\!\left[\begin{array}{c}
\mathbb{E}[X]\\
\mathbb{E}[X^2]
\end{array}\right]
+
\left[\begin{array}{ccc}
D&0&0\\
0&D&0\\
0&0&D
\end{array}\right]\!\!\left[\begin{array}{c}
\mathbb{E}[K_1]\\
\mathbb{E}[K_1^2]\\
\mathbb{E}[K_1K_2]
\end{array}
\right]\nonumber\\
&
\!\!+\!\!
\begin{bmatrix}
0&0&-1&0&0&0\\
2K_3&-2K_3&0&0&-1&0\\
0&0&2K_3&-2K_3&0&-1
\end{bmatrix}
\!\!\!
\begin{bmatrix}
\mathbb{E}[XK_1]\\
\mathbb{E}[X^2 K_1]\\
\mathbb{E}[XK_2]\\
\mathbb{E}[X^2 K_2]\\
\mathbb{E}[XK_1 K_2]\\
\mathbb{E}[XK_2^2]\\
 \end{bmatrix}
\label{example-momeq}
\end{align}
which is obtained from eq. (\ref{parmomeq}) with $\bm{\hat{X}}^{\bm{\zeta}} = X^{\zeta_1} K_1^{\zeta_2} K_2^{\zeta_3}$. 
For later convenience, we define 
\begin{align}
\rho := \max \zeta_1, \ \sigma := \max (\zeta_2 + \zeta_3), 
\end{align}
which are the parameters that determine the order of moments included in the truncated moment equation. These parameters are used to control the tradeoff between accuracy and computation time. For example, eq. (\ref{example-momeq}) corresponds to the case of $\rho = \sigma = 1$.
Substituting %
eq. (\ref{mom_gamma}) into the corresponding entries in eq. (\ref{example-momeq}), we obtain the linear equation of the optimization problem in Proposition 1.

\par
We further narrow the solution space of eq. (\ref{example-momeq}) by using $\bar{H}_0, \bar{H}_1, \bar{H}_2$, and $\bar{H}_3$. 
For $\rho = \sigma = 1$, we use 
\begin{align}
\bm{g}_0&=[1,X,K_1,XK_1,K_2,XK_2]^\top, \nonumber\\
\bm{g}_1&=\bm{g}_2=[1,X]^\top,~~\bm{g}_3=[1,K_1,K_2]^\top,
\notag 
\end{align}
and 
$c_1(\bm{K}) = K_1, c_2(\bm{K}) = K_2,\ d_1(X) = X$, which allows for including all moments in eq. (\ref{example-momeq}) into the matrices.

\begin{figure}[t]
\centering
\includegraphics[width=8.5cm]{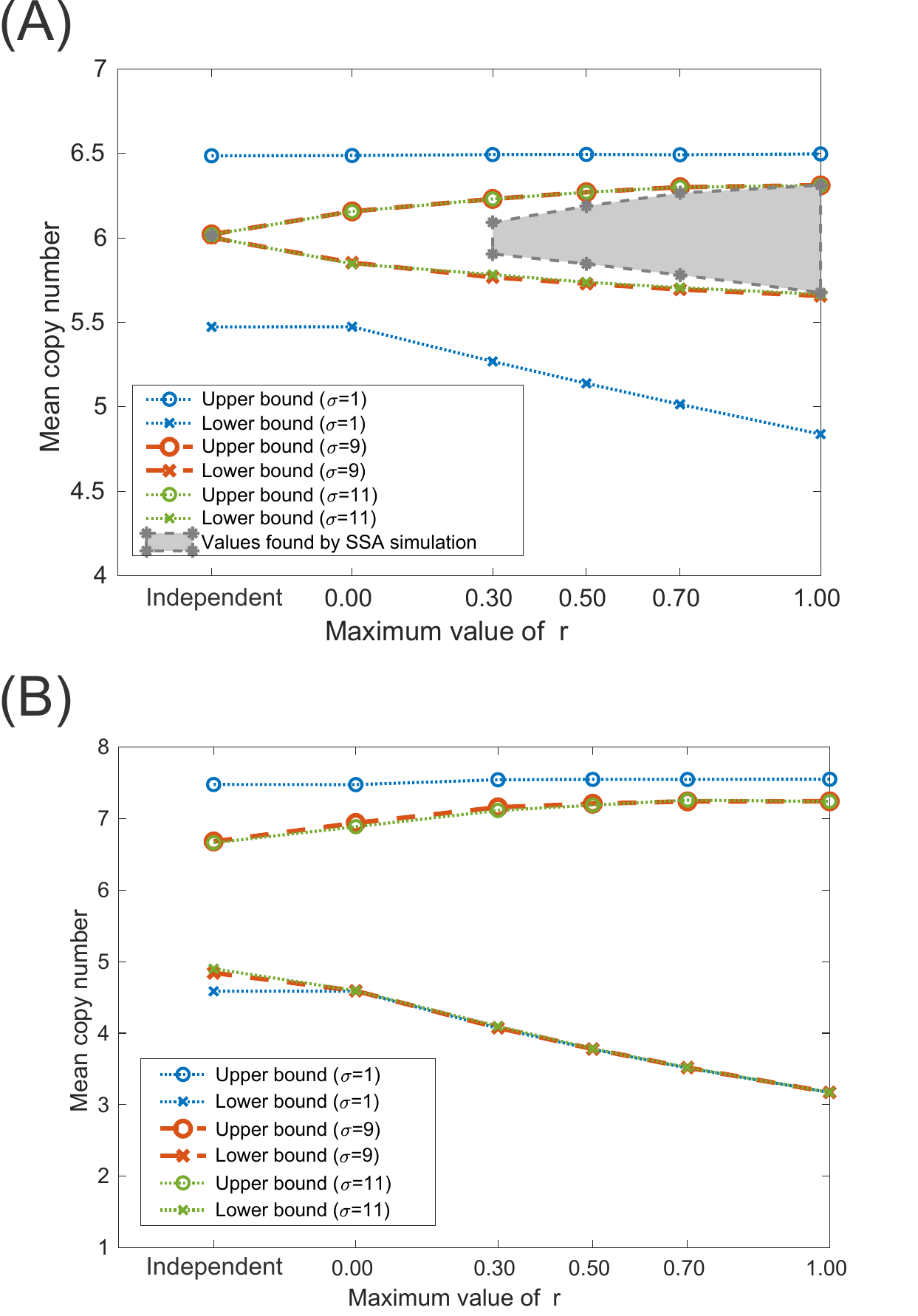}
\caption{Bounds of maximum and minimum values of mean copy number $\mathbb{E}[X]$ for the reaction system (\ref{ModelMonomerDimer}). (A) gamma distribution is assumed for $\mathbb{P}_{K_1}$ and $\mathbb{P}_{K_2}$. (B) Only the  first and the second order moments are assumed for  $\mathbb{P}_{K_1}$ and $\mathbb{P}_{K_2}$.}
\label{Range_Mean_NL}
\end{figure}

Finally, based on Proposition 1, we compute mathematically valid upper and lower bounds of the mean copy number $\mathbb{E}[X]$ for different values of $r$.  
The results are illustrated in Fig. \ref{Range_Mean_NL}(A), where $\rho = 5$ is used.
We observe from Fig. \ref{Range_Mean_NL}(A) that, for each $r$, the gap between the upper and lower bounds monotonically decreases with increasing $\sigma$. This is because the constraints for the optimization program with smaller $\sigma$ become a subset of those for the larger ones. This feature allows us to automate the choice of $\sigma$ by iteratively solving the optimization problem by increasing $\sigma$ until 
the gap between the bounds reaches to an acceptable range or the decrease of the gap starts stalling. %
Another important observation is that the gap increases monotonically with $r$. This gap shows the uncertainty of the mean copy number that originates from the uncertainty of the joint distribution $\mathbb{P}_K$ due to the possible correlation of the parameters. In other words, Fig. \ref{Range_Mean_NL}(A) shows the worst-case value of the uncertain moment for each $r$. %
The gap becomes zero (with two significant digits) for $\sigma=9$ when the parameters are independent, for which case $\mathbb{P}_K$ is uniquely determined as the product of the two gamma distributions. %

\par
This result can be verified by generating sample paths for each $r$ and plotting the intervals of the mean copy number as shown in Fig. \ref{Range_Mean_NL}(A). 
More specifically, the sample paths are generated by the following procedure: 
\begin{enumerate}%
 \item  Generate 100000 random numbers $K_{1,i}$ and $K_{2,i}\ (i=1,2,\ldots,100000)$ from the gamma distributions $\mathbb{P}_{K_1}$ and $\mathbb{P}_{K_2}$, respectively.
\item Sort each parameter in ascending order and make pairs of parameters $(K_{1,i_1}, K_{2,i_2})$. 
\item Randomly choose two pairs of parameters, say $(K_{1,i_1}, K_{2,i_2})$ and $(K_{1,j_1}, K_{2,j_2})$, and swap $K_2$ so that 
$(K_{1,i_1}, K_{2,j_2})$ and $(K_{1,j_1}, K_{2,i_2})$. 
\item Repeat (2) and (3) unless $\mathrm{corr}(K_1, K_2) \le r$ 
\item Run the stochastic simulation algorithm (SSA) \citep{Gillespie1976} for each parameter pair, and record the mean copy number at time $t=1440$.
\end{enumerate}
In short, positively correlated parameter pairs are generated by step (2), and then the correlation is reduced in step (3). To make negatively correlated pairs, 
\begin{enumerate}
\item[(6)] Run (1)-(5) again, but $K_2$ is sorted in descending order in step (2) and $\mathrm{corr}(K_1, K_2) \ge -r$ in step (4). 
\item[(7)] Plot the range of the mean copy numbers obtained in steps (5) and (6).
\end{enumerate}

The average computation time for generating a single sample path was 0.6094 second for $r = 0.30, 0.50, 0.70$, and 1.00 in Fig. \ref{Range_Mean_NL}(A). We observed that 10000 or more sample paths were necessary to obtain informative asymptotic bounds, which equates 6094 second in average for each bound. On the other hand, the computation time of the proposed optimization was 2400 second in average.

In general, obtaining asymptotic bounds using the sample path generation approach tends to be prohibitively hard when there are fewer assumptions on the parameter distributions. For example, if we remove the assumption that the marginal distributions $\mathbb{P}_{K_1}$ and $\mathbb{P}_{K_2}$ are gamma distributions, there are many other possible distributions satisfying the constraints of the mean, the variance, and the correlation $|\mathrm{corr}(K_1, K_2)| \le r$. Consequently, searching for all possible distributions would be very hard  by the Monte Carlo approach. %
On the other hand, the computational cost of the proposed optimization remains almost the same even for such cases since the bounds can be computed simply by changing the constraints of the optimization. %

\subsection{Analysis for fully unknown parameter distributions}
\label{complete-example}
Next, we consider a more practical scenario where the marginal distributions of the parameters $K_1$ and $K_2$ are not completely known, but only their first and second order moments are. %
We assume $K_3 = 0.02$ and $D = 5$, and the correlation between the two parameters is assumed to be $|\mathrm{corr}(K_1, K_2)| \le r$, which is the same as the previous example. 
Thus, the only difference from the previous example is that the parameter distribution $\mathbb{P}_K$ has larger uncertainty in that the marginal distribution is not unique. 

\par
We formulate the same optimization problem as the previous example in Section \ref{partial-example} except that the third and the higher order moments for $K_1$ and $K_2$, {\it i.e.,} $\mathbb{E}[K_1^r]$ and $\mathbb{E}[K_2^r]$ with $r \ge 3$, are set as variables. 
The mathematically valid upper and lower bounds of the mean copy number computed by the optimization program are plotted in Fig. \ref{Range_Mean_NL}(B).  
As expected, the gap between the upper and the lower bounds becomes larger than those in Fig. \ref{Range_Mean_NL}(A) since the parameter distribution $\mathbb{P}_K$ in this example has larger uncertainty than in the previous example.  
It should be noted that the marginal distributions of the parameters are no longer limited to the gamma distributions. Since parameterization of such uncertain parameter distributions is not available, it is prohibitively difficult to obtain a reasonable estimation of valid bounds by using the sampling based approach such as the SSA \citep{Gillespie1976}. 
The proposed approach, on the other hand, gives mathematically valid bounds, and thus, it is useful for rational engineering and analysis of stochastic chemical reaction systems. 

\par
\noindent
{\bf Remark 3.}
All optimization problems were solved with SeDuMi 1.3.2 \citep{Sturm1999} on MATLAB 2021a.
To avoid numerical instability of the solver, the variables were normalized by constants as shown in Appendix \ref{normalization-sec}.

\section{Conclusion}
\label{concl-sec}
We have proposed a computational framework to analyze the worst-case stationary moments of the molecular copy number distributions in stochastic chemical reactions with parametric uncertainty. Specifically, a mathematical optimization method has been developed to compute the intervals of the possible moment values of uncertain moment equations whose parameters are given only partially using the statistics of the parameter distributions. 
A distinctive feature of the proposed method is that it has been derived without approximating the governing equation of the stochastic chemical reactions, {\it i.e.}, the CME, unlike many other approaches reviewed in Section \ref{intro-sec}.  In other words, the moments of interest are guaranteed to be within the computed bounds for all possible parameter distributions satisfying the given statistics. This feature is useful for model-based rational engineering of biomolecular circuits, where the robustness of synthetic reactions is important.
\noindent
{\bf Acknowledgments:} 
This work was supported in part by JSPS KAKENHI Grant Number JP16H07175, JP18H01464, and 21H01355. %
\bibliographystyle{agsm}
\bibliography{manuscript}

\appendix

\section{Scaling of the variables for the optimization}
\label{normalization-sec}
Scaling constants were introduced to avoid numerical instability caused by rounding error.
Let
\begin{align}
& \mathbb{E}_{\rm scaled} [X^{\alpha} K_1^{\beta_1} K_2^{\beta_2}]  := 
\frac{1}{C(\alpha, \beta_1, \beta_2)} \mathbb{E}[X^{\alpha} K_1^{\beta_1} K_2^{\beta_2}] \notag 
\end{align}
denote a scaled moment, where
$C({\alpha},\beta_1, \beta_2):=C_{X}^{\alpha} C_{K_1}^{\beta_1} C_{K_2}^{\beta_2}$ %
with $C_{X}$ and $C_{K_i}$ being given constants associated with the copy number $X$ and the parameters $K_i~(i=1,2)$. %

\par
When solving the optimization problem, we reformulated equivalent constraints by dividing the stationary moment equation (\ref{parmomeq}) with $\bm{\zeta} = [\alpha, \beta_1, \beta_2]^\top$ by $C(\alpha, \beta_1, \beta_2)$.
The variables in $\bar{H}_i$ were also replaced with the scaled moments with appropriate scaling. The constants were $C_{X} = 5, C_{K_1} = 3$ and $C_{K_2} = 0.7$ for Fig. \ref{Range_Mean_NL}.

\end{document}